\documentclass{article}

\usepackage{arxiv}

\usepackage[utf8]{inputenc} 
\usepackage[T1]{fontenc}    
\usepackage{hyperref}       
\usepackage{url}            
\usepackage{booktabs}       
\usepackage{amsfonts}       
\usepackage{nicefrac}       
\usepackage{microtype}      
\usepackage{lipsum}		
\usepackage{graphicx}
\usepackage{natbib}
\usepackage{doi}

\usepackage{amsmath}
\usepackage{algpseudocode}
\usepackage{algorithm}

\usepackage{verbatim}

\title{A Custom IC Layout Generation Engine Based on Dynamic Templates and Grids}

\date{July 16, 2022}	

\author{Taeho Shin\\
        Department of Electronic Engineering\\
        Hanyang University\\
        Seoul, Korea\\
        \texttt{sth4101@hanyang.ac.kr} \\
	\And
        Dongjun Lee\\
        Department of Electronic Engineering\\
        Hanyang University\\
        Seoul, Korea\\
        \texttt{lreedong12@hanyang.ac.kr} \\
        \And
        Dongwhee Kim\\
        Department of Electronic Engineering\\
        Hanyang University\\
        Seoul, Korea\\
        \texttt{dwkim565@hanyang.ac.kr} \\
        \And
        Gaeryun Sung\\
        Department of Electronic Engineering\\
        Hanyang University\\
        Seoul, Korea\\
        \texttt{sungkr3209@hanyang.ac.kr} \\
        \And
        Wookjin Shin\\
        Department of Nano-scale Semiconductor Engineering\\
        Hanyang University\\
        Seoul, Korea\\
        \texttt{dksmlthakd01@hanyang.ac.kr}\\
        \And
        Yunseong Jo\\
        Department of Electronic Engineering\\
        Hanyang University\\
        Seoul, Korea\\
        \texttt{jhj101777@hanyang.ac.kr}
        \And
        Hyungjoo Park\\
        Department of Electronic Engineering\\
        Hanyang University\\
        Seoul, Korea\\
        \texttt{pikkoro97@hanyang.ac.kr}
        \And
        Jaeduk Han\\
        Department of Electronic Engineering\\
        Hanyang University\\
        Seoul, Korea\\
        \texttt{jdhan@hanyang.ac.kr}
}

\renewcommand{\shorttitle}{A Custom IC Layout Generation Engine Based on Dynamic Templates and Grids}

\hypersetup{
pdftitle={A Custom Layout Generation Engine Based on Dynamic Templates and Grids for Advanced CMOS Integrated Circuits},
pdfsubject={Integrated Circuits},
pdfauthor={Jaeduk Han},
pdfkeywords={Full-custom circuits, Analog circuits, Layout generation, Design rules},
}

\begin{document}
\algnewcommand\algorithmicswitch{\textbf{switch}}
\algnewcommand\algorithmiccase{\textbf{case}}
\algnewcommand\algorithmicassert{\texttt{assert}}
\algnewcommand\Assert[1]{\State \algorithmicassert(#1)}%
\algnewcommand\algorithmicforeach{\textbf{for each}}
\algdef{S}[FOR]{ForEach}[1]{\algorithmicforeach\ #1\ \algorithmicdo}

\maketitle

\begin{abstract}
This paper presents an automatic layout generation framework in advanced CMOS technologies. The framework extends the template-and-grid-based layout generation methodology with the following additional techniques applied to produce optimal layouts more effectively. First, layout templates and grids are dynamically created and adjusted during runtime to serve various structural, functional, and design requirements. Virtual instances support the dynamic template-and-grid-based layout generation process. The framework also implements various post-processing functions to handle process-specific requirements efficiently. The post-processing functions include cut/dummy pattern generation and multiple-patterning adjustment. The generator description capability is enhanced with circular grid indexing/slicing and conditional conversion operators. The layout generation framework is applied to various design examples and generates DRC/LVS clean layouts automatically in multiple CMOS technologies.
\end{abstract}

\keywords{Full-custom circuits, Analog circuits, Layout generation, Design rules}

\section{Introduction}
Owing to the advances in semiconductor technology, the computing performance of integrated circuits has been enhanced dramatically, enabling a number of data-centric applications. The evolution continuously pushes the demands on the high-performance computing chips fabricated in deeply-scaled CMOS technologies. Recently, the nominal channel length of transistors for the advanced CMOS technologies has been reduced down to 3-5 nm ranges (\cite{2_irds}). Although the device scaling trend has contributed to enhancing the performance of the integrated circuits, it also results in complications of transistor structures and characteristics (\cite{2_irds,1_han,3_loke}). Therefore, significant resources have been spent on building high-performance integrated circuits in recent technology nodes. Unfortunately, it is expected that the device complication will get worse as new device structures such as Gate-All-Around FET (GAAFET) in \cite{4_huang, 5_smith} or Tunneling FET in \cite{6_krishnamohan} are considered for the next primitive device elements in combination with complex local interconnects and multiple patterning technologies for dense routing.

In order to address the design complication and cost increase issues in advanced semiconductor technologies, this paper describes a layout design automation tool for nanometer CMOS technologies. The framework adopts various design and automation techniques (such as dynamic template and grid and post-processing capabilities) to produce compact and high-quality layouts automatically and reduce the design cost.

\subsection{Prior works on layout generation}
Techniques to automate the process of physical layout generation have been investigated for decades. The previously reported generation methods generally fall into two categories: rule-based (\cite{7_alsyn,8_koan,9_iprail}) and template-based approaches (\cite{1_han,10_lopez,11_lopez,12_iip,13_bag2}). The rule-based methods utilize design-rule parameters to compute the coordinates of polygons for the structures of devices and routing wires. The rule-based methods can be seamlessly applied to new technologies or device flavors simply by applying a new set of design rule parameters corresponding to the porting configuration. However, the rule-based approaches basically assume identical device structures and/or layer organizations, which do not hold anymore in advanced CMOS technologies where the device structures vary significantly across nodes. Another challenge is the increased overhead of interpreting complex design rules in modern technologies. 

The template-based approach is more suitable to handle design-rule complexities and technology differences. In the template-based layout generation method, the device/circuit templates predefine the structures of physical components, and the layout generation is performed by placing the templates and routing grids on technology-specific grids. In terms of the specific ways of defining templates, \cite{10_lopez,11_lopez} use parameterized template cells while \cite{1_han} uses handcrafted templates for structural optimization of primitive devices. \cite{12_iip} focuses on building design environments to foster reuse by organizing design-specific generators in a holistic framework, while the design entries are abstracted in higher levels rather than in device templates. \cite{14_align,15_align2} combine the rule-based and template-based approaches to produce the templates and grids from a basic set of design rules with additional automated placement and routing capabilities for further productivity enhancements. Recent researches utilize machine-learning techniques (\cite{15_align2,16_magical}) or open-source toolchains (\cite{17_openram,18_openroad,19_openlane}) to enhance design productivity in CMOS logic technologies with template-based layout generation methods.

\subsection{Contributions and Organization of This Work}
\begin{figure}[h]
  \centering
  \includegraphics[width=0.3\linewidth]{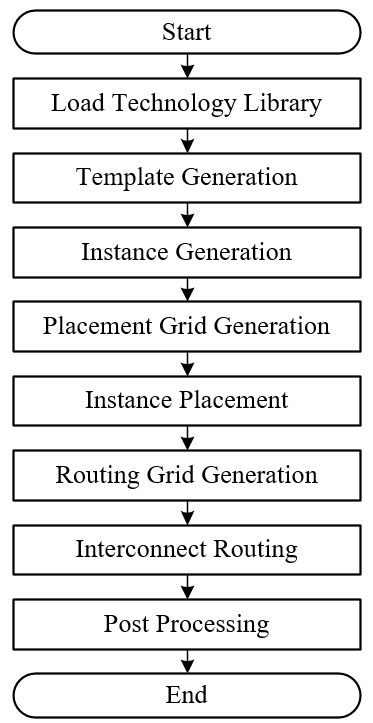}
  \caption{Proposed layout generation process.}
  \label{fig1}
\end{figure}

This paper introduces an automated layout generation engine developed to meet various requirements of advanced CMOS technologies. It extends the templates-and-grid-based layout generation method to dynamically generate the templates and grids, and produce the optimal device and routing patterns. To be specific, the templates are reconfigured and regenerated during the layout generation phase, producing a group of instances and geometries, and encapsulating them as a virtual instance for target design intents and process requirements. The placement and routing grids are configured dynamically as well, which enhances the area efficiency and flexibility. In addition to that, various post-processing tasks are implemented to address process-specific needs in advanced technologies, which include cut pattern generation and dynamic coloring. The overall layout generation flow is adjusted to support the proposed functions and is illustrated in Figure \ref{fig1}. In particular, the routing grid generation step is inserted after the (generated) instance placement step to consider the placement information for optimal grid generation. Various generator description techniques are introduced as well to support the aforementioned new features in the framework. Virtual instances are introduced to encapsulate a group of layout objects as a single instance object. Circular mapping and conditional indexing techniques are used to index and slice array objects and grids, which enhance the description efficiency. 

The organization of the paper is as follows: First, major features and related operating principles are explained in Section 2. After that, Section 3 shows various layout generation examples using the framework in advanced CMOS technologies and concludes this paper.

\section{Major features to implement automatic layout generations in advanced CMOS technologies}
\subsection{Dynamic templates and virtual instances to achieve high area efficiency as well as productivity}
The most critical requirement on layout generation methodologies is the encapsulation of design rules, of which complexity increases significantly as the process scales down. The most common approach to abstract the design rules is the use of templates and grids (\cite{1_han,10_lopez,11_lopez,12_iip,13_bag2,14_align,15_align2}). The templates implement primitive physical elements such as transistors, resistors, and capacitors. During the layout generation phase, each template is one-to-one mapped to a process-specific layout instance in the target technology, and the layout instances are then placed on dedicated placement grids, followed by wire-routing operations to provide connectivity between the placed instances. This template-and-grid-based layout generation method effectively hides the detail of internal structures of the layout instances and interconnects in upper levels, enhancing design efficiency and process portability. The methodology also enables parameterized layout generation by creating the instances in arrayed forms and configuring the dimension of the array instances.

\begin{figure*}[h]
  \centering
  \includegraphics[width=1.0\columnwidth]{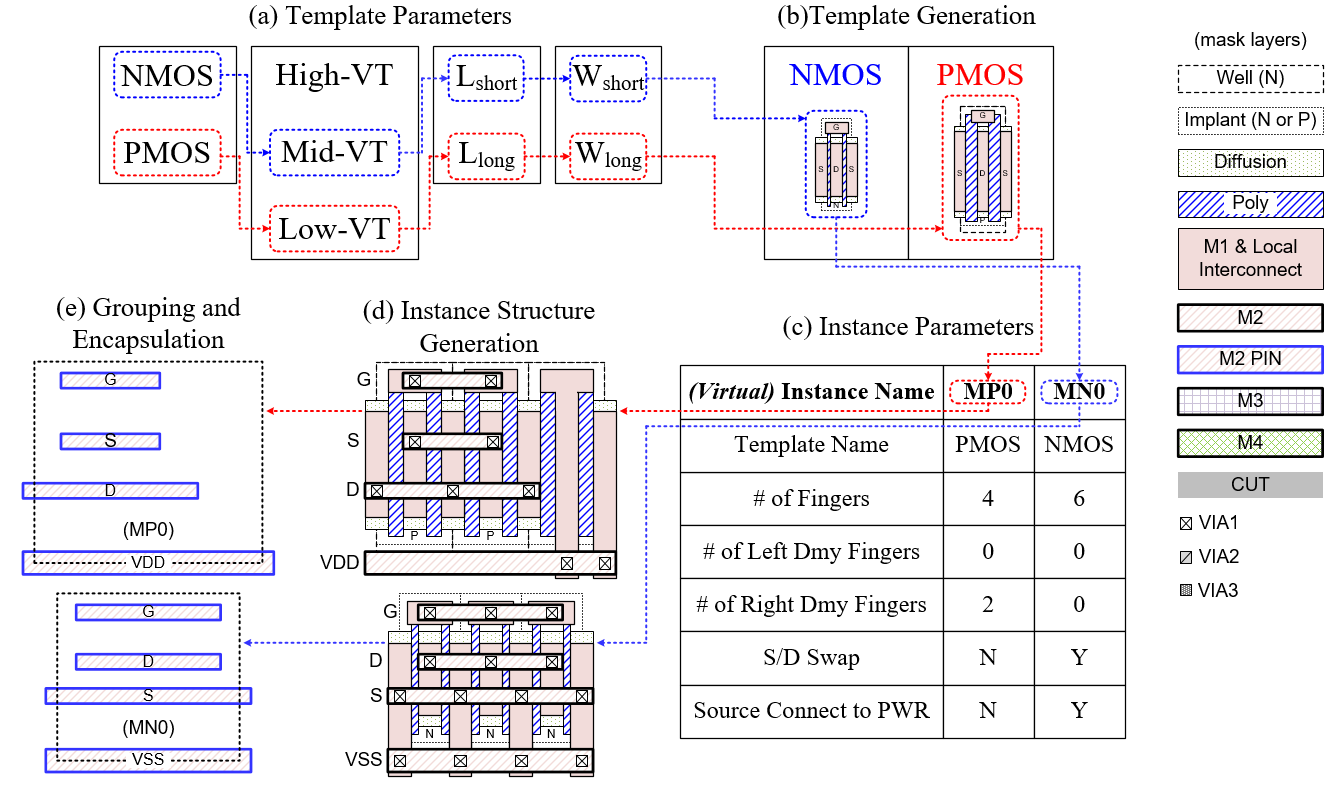}
  \caption{Dynamic template and instance generation process.}
  \label{fig2}
\end{figure*}

The proposed framework improves the instance-mapping templates by (re)organizing the templates based on design/technology-specific requirements. This technique is called dynamic template generation, as the templates are dynamically mapped to a group of (re)organized layout structures during run time (Figure \ref{fig2}), rather than simply mapped to static instances. In the previous static template approach (\cite{1_han}), as the templates are simply mapped to single instances in each technology, their flexibility and parameterization capabilities are constrained by the accessibility and variety of the handcrafted instances. For example, static templates require a new set of primitive device layouts for each variant (short/long channel, different threshold voltage (VTH) flavors, etc.), which potentially causes fragmentation issues. For example, if a process provides three different VTH options (e.g. standard-VT, low-VT, high-VT), there should be at least three transistor templates if the templates are simply mapped to static instances. On the other hand, the proposed dynamic template generation technique performs the instantiation process by creating a layout group composed of multiple layout objects (instead of single instances) based on its design and technology parameters. Therefore, the dynamic templates greatly enhance the coverage and flexibility of layout generators by describing a set of devices with a single template code and (re)generating actual templates based on the input parameters. 

The ‘dynamically-generated’ templates then produce actual component instances (e.g., transistors) by placing the generated layout object group in an array and creating peripheral structures (such as dummies, routing structures, internal probes, and power rails/plugs). Therefore, it is convenient to abstract the layout object group as a single instance with its bounding box and terminal coordinates. The proposed framework implements a wrapper class called $VirtualInstance$ for the purpose (Figure \ref{fig2}(e)). The virtual instance is an abstraction of a layout object group composed of its sub-elements. The organization, physical shape, and relative locations of the sub-elements are determined by the type and parameters of their master virtual instance. As the virtual instance object is handled in the same way as a static instance, it offers the same user experience as native instances during the layout generator description phase. Designers simply provide the type, location, and design parameters of the virtual wrapper instance, and the placement engine in the proposed framework computes physical parameters and coordinates of sub-elements during the generation phase. In particular, the placement coordinate $p_i$ of the \textit{i-th} sub-element of a virtual instance is expressed as follows: 

\begin{equation}
  \vec{p_i}=\vec{x}+0.5\cdot (I-T)\cdot \vec{s}+T_i\cdot \vec{x_i}
\end{equation}

where $\vec{x}$ is the vector representing the origin of the virtual instance, and $T$ is the transform matrix of the virtual instance, which represents the rotational and mirroring information of the instance (\cite{1_han}). The value of $T$ for the virtual instance’s transform parameter is summarized in Table 1. $T_i$ is the transform matrix of the sub-element, and $\vec{x_i}$ is the positional vector of the sub-element from the origin of the virtual instance.
The framework converts all relevant coordinates based on the expression (1) and places all sub-elements of the virtual instance. The main benefit of using virtual instances is that users can treat the generated virtual instance (which is composed of multiple sub-elements) as a single nominal instance for their placement, accessing coordinates, and array creation with richer options to produce a variety of layout geometries and structures. This encapsulation capability greatly boosts the reusability of generators regardless of their application technology, from legacy planar processes to state-of-the-art technologies such as FinFET.

\begin{table}
  \caption{Example of Transform Matrices for Placement Of Virtual Instances}
  \centering
  \label{tab:matrix}
  \begin{tabular}{cccl}
    \toprule
    Type & Matrix $T$ & $0.5\cdot (I-T)$ & Description\\
    \midrule
    R0 & $\begin{bmatrix} 1 & 0 \\ 0 & 1 \end{bmatrix} $
       & $\begin{bmatrix} 0 & 0 \\ 0 & 0 \end{bmatrix} $& Place without transformation \\
       \\
    MX & $\begin{bmatrix} 1 & 0 \\ 0 & -1 \end{bmatrix} $
       & $\begin{bmatrix} 0 & 0 \\ 0 & 1 \end{bmatrix} $ & Mirror on x-axis \\
       \\
    MY & $\begin{bmatrix} -1 & 0 \\ 0 & 1 \end{bmatrix} $
       & $\begin{bmatrix} 1 & 0 \\ 0 & 0 \end{bmatrix} $ & Mirror on y-axis \\
       \\
    R180 & $\begin{bmatrix} -1 & 0 \\ 0 & -1 \end{bmatrix} $
         & $\begin{bmatrix} 1 & 0 \\ 0 & 1 \end{bmatrix} $ & Rotate by 180 degrees \\
  \bottomrule
\end{tabular}
\end{table}

\subsection{Dynamic routing grid generation for design-specific interconnects}
\begin{figure}[h]
  \centering
  \includegraphics[width=0.5\linewidth]{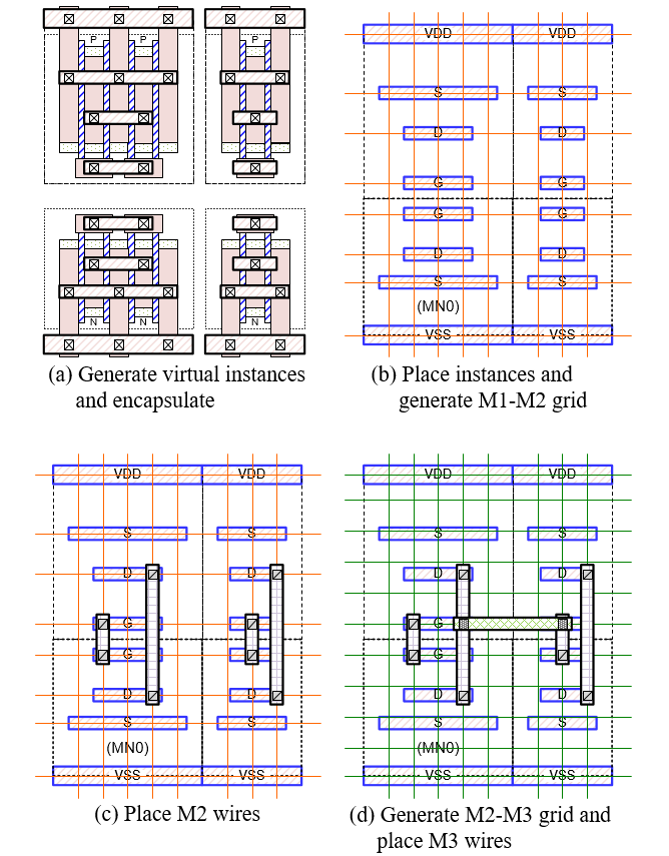}
  \caption{Dynamic generation of routing grids.}
  \label{fig4}
\end{figure}

After generating layout structures either by typical instances or virtual instances, the structures are placed, and wires are routed on technology-specific grids to finish the layout generation process. However, the area consumption may increase due to the introduction of placement and routing grids, especially when uniform grids are used without careful consideration of area efficiency. In fact, it is difficult to choose the optimal grid configuration without utilizing the type of devices to be placed and their placement information. This area/routing-track inefficiency is one of the main reasons that the grid-based routing method has not been adopted widely for custom layout designs in legacy technologies where designers have more freedom on device placement and routing. In advanced technologies, while the area penalty is not as significant as in legacy nodes due to the restricted design rules and placement/routing options, optimizing routing grids for design-specific structures and requirements yields further enhancements in performance and area consumption. In particular, the dynamic template generation approach requires online adjustment of routing grids for the generated templates. Therefore, the framework improves the predefined and static grids or pin-based grids (\cite{20_dram}) in previous works and supports the dynamic creation of placement/routing grids as a form of generators so that designers can reproduce the grids flexibly for their specific requirements or design structures (Figure \ref{fig4}). In other words, the grid generator receives the design requirements as its input parameters and produces grids based on the constraints. This approach enables designers to choose optimal routing grids for design-specific requirements, such as compact routing for high routing density or thick routing for high current flowing capability. 

\subsection{Post-processing operations and related algorithms to deal with various processes and requirements}
There are various process-specific tasks that need to be carried out after the vanilla placement and routing operations in advanced CMOS technologies, especially to meet design rules. For example, when wires are densely routed in a small area, cut patterns may be placed to meet stringent spacing rules. This is because when multiple wires are routed close to each other, they are fabricated by producing a single aggregated metal pattern and cutting it into several parts (\cite{21_vladimir}). Other examples include extending wires to meet metal area rules, inserting surrounding dummy patterns, and adjusting coloring (multiple patterning) parameters for pattern alignment. These process-specific tasks are hard to be integrated into the basic placement and routing procedures, as they are supposed to be executed homogeneously regardless of the process technology. Therefore, the proposed framework implements dedicated post-processing steps after the basic layout generation process in order to perform the process-specific operations effectively. 

One of the most representative process-specific tasks is the cut-pattern generation introduced in the previous subsection. As there is a growing demand for automated and hierarchical generation of cut patterns as the routing dimension scales down continuously (\cite{21_vladimir}), the framework implements a simple and efficient post-processing algorithm to create cut patterns for routing patterns hierarchically as follows: First, the locations and spacing parameters of routing patterns are extracted. After that, scanning the entire routing elements, for each element, check if the element is apart from other routing elements with enough spacing. If a routing pair does not meet the spacing rule, a cut pattern is inserted between the two routing elements to resolve related design rule violations. After inserting the cut patterns between routing wires, additional cuts are selectively placed at the edges of wires located at boundaries, as the boundary routing elements can produce spacing rule violations by interacting with neighboring routing elements in upper levels. The boundary cuts are not generated for pin routing wires, as the pin locations are captured, and cut patterns can be generated in upper levels. The pseudo-code in Algorithm \ref{alg:cutgen} implements the proposed cut insertion algorithm:

\begin{algorithm}
  \centering
  \caption{Cut Pattern Generation}
  \label{alg:cutgen}
  \begin{algorithmic}[1]
  \State {R is the group of entire routing wires and instance ports with the layer of interest.}
  \State {\Call{Violate\_Spacing\_Rule}{$r1, r2$}: Returns true if r1 and r2 violates the spacing rule between them.}
  \State {\Call{Located\_Lend}{$r, R$}: Returns true if r is located at the left-most side of routes in the same track in the routing group R.}
  \State {\Call{Located\_Rend}{$r, R$}: Returns true if r is located at the right-most side of routes in the same track in the routing group R.}
  \State
  \Procedure{Cut\_Pattern\_Gen}{$r1,~r2$}
    \ForEach{$r1 \in R$}
      \ForEach{$r2 \in R$}
        \If{\Call{Violate\_Spacing\_Rule}{$r1, r2$} is $true$} 
          \State Create a cut pattern between r1 and r2
        \EndIf
      \EndFor
      \If{\Call{Located\_Lend}{$r1, R$} is $true$ and r1 is not a pin} 
        \State Create a cut pattern at the left side of r1
      \EndIf
      \If{\Call{Located\_Rend}{$r1, R$} is $true$ and r1 is not a pin} 
        \State Create a cut pattern at the right side of r1
      \EndIf
    \EndFor
  \EndProcedure
  \end{algorithmic}
\end{algorithm}

\begin{figure}[h]
  \centering
  \includegraphics[width=0.4\linewidth]{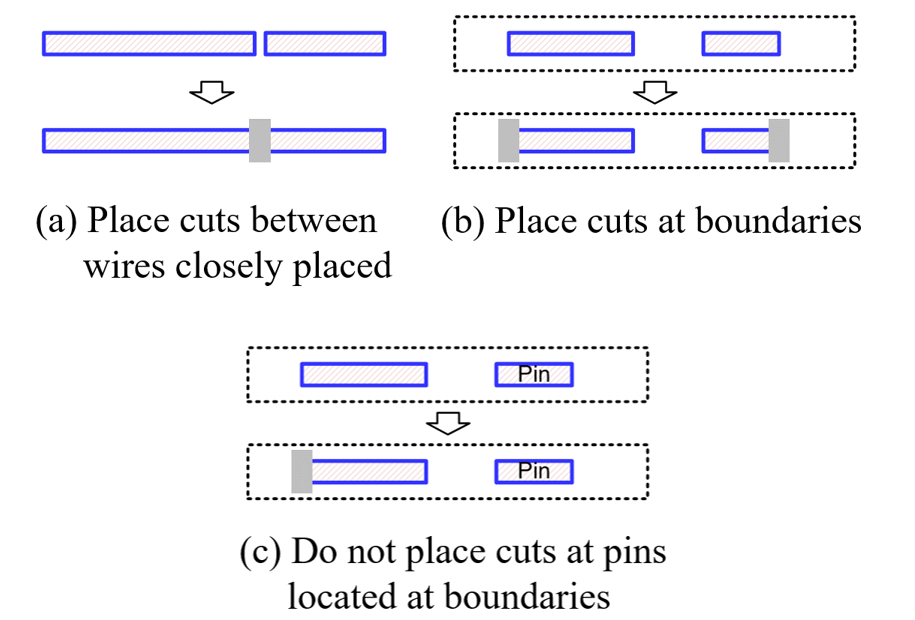}
  \caption{Exemplar cut patterns generated from Algorithm 1.}
  \label{fig5}
\end{figure}

Layout examples corresponding to the cut-pattern generation cases are illustrated in Figure \ref{fig5}. As the cut-pattern generation step is required only in advanced technologies, it is implemented as a post-process function so that its execution is determined based on the target technology node.

\subsection{Advanced grid accessing and conversion functions based on cyclic indexing and conditional operators}
The description capability of the generator scripts is critical for implementing highly customized and flexible layout generators efficiently. In particular, users should be able to readily access grid elements for given constraints and convert the coordinate values across various grid systems. The framework inherits the basic grid concepts proposed in previous works (\cite{1_han,14_align}) with additional features such as 1) cyclic data structures and grid access functions for advanced slicing/indexing (\cite{22_oliphant}) and 2) conditional operators (\cite{23_pandas}) for reverse mapping. 

As the grid and array parameters expand over the entire physical space in a periodic manner, data structures for representing the grid parameters should support rotational indexing. The proposed layout generation framework implements dedicated two data structures for this purpose, $CircularMapping$ (for 1-D lists) and $CircularMappingArray$ (for 2-D arrays), to describe the cyclic behavior. To be specific, for a one-dimensional cyclic list with r elements, the element corresponding to the index i is accessed by the following expression:

\begin{align}
e_i = \begin{cases}
e_{i\%r} & \text{if $ i\geq 0$} \\
e_{r-i\%r} & \text{otherwise}
\end{cases}
\end{align} 

\begin{figure}[h]
  \centering
  \includegraphics[width=0.5\linewidth]{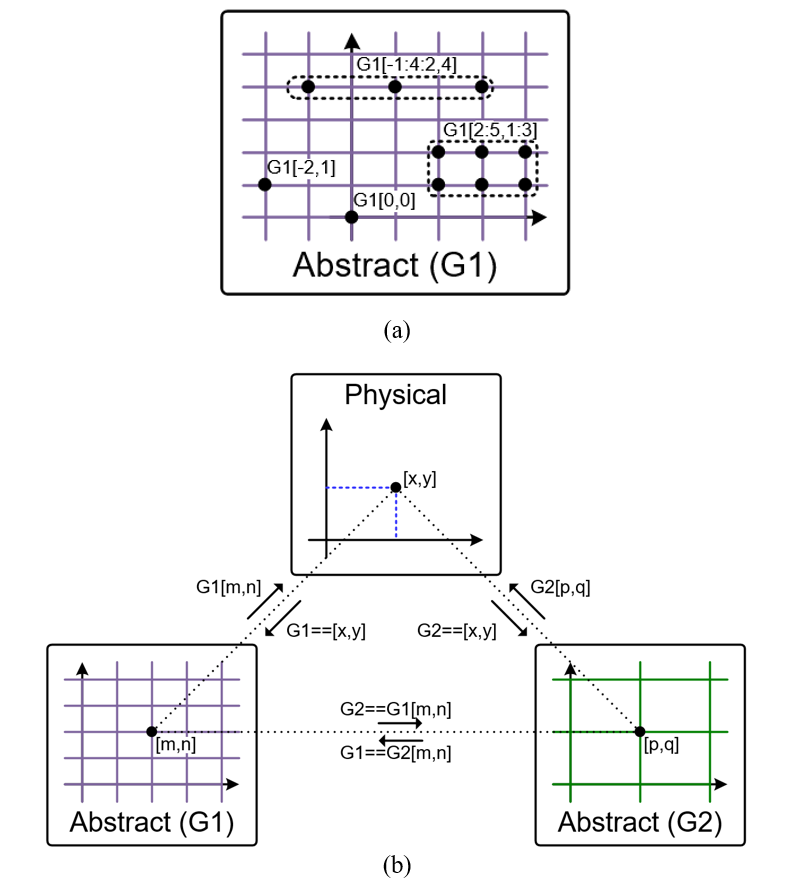}
  \caption{(a) Grid slicing and accessing functions and (b) Grid accessing by conditional operators.}
  \label{fig6}
\end{figure}

The two-dimensional circular mapping array is implemented by combining two one-dimensional lists. In addition to the cyclic array, the indexing capability is enhanced further by 1) advanced slicing and indexing functions for grid access and 2) conditional operators for grid conversions. The slicing and indexing functions were originally implemented in \cite{22_oliphant} for scientific computing and adopted in \cite{1_han} to access and manipulate array objects (instances). In this work, the slicing and indexing functions are utilized for the grid indexing as well to access multiple coordinates efficiently and implement complex access functions such as finding the overlapping region of two objects. In addition to that, conditional operators in \cite{23_pandas} are used for implementing grid conversion functions for better generator description capabilities by describing conditions for the coordinates to access with conditional and logical operators. To be specific, while the abstract to physical coordinate conversion is done by the normal indexing operator, the reverse mapping is supported by the conditional operator. Figure \ref{fig6} illustrates the examples of grid slicing and indexing (Figure \ref{fig6}(a)) and grid conversion by conditional operators (Figure \ref{fig6}(b)). Due to the grid handling features, users can describe the various placement and routing structures efficiently. 

\section{Layout Generation Examples}
\begin{figure*}[h]
  \centering
  \includegraphics[width=1.0\columnwidth]{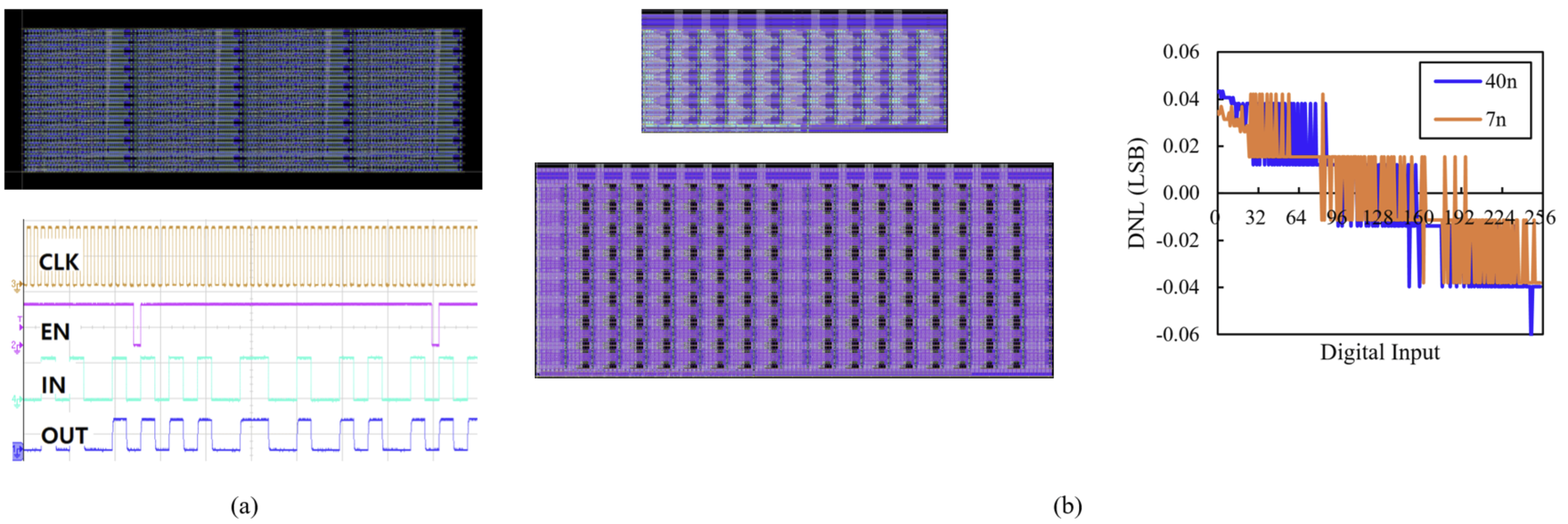}
  \caption{Generated layout examples: (a) Layout and measured signal waveforms for a generated custom scan for analog circuit conditioning, (b) Layout of current DACs with variable bit width and transistor sizing parameters in 40 nm, and simulated DNLs in 40 nm and 7 nm technologies.}
  \label{fig7}
\end{figure*}

In order to demonstrate the parameterized and flexible layout generation capability of the proposed framework, various physical designs are produced in different technologies. First of all, Fig. 7(a) shows a generated layout of a custom scan cell for reading out and writing digital bits across different voltage and/or chip domains. The custom scan generator is capable of programming the number of bits, read/write configurations, and optional level-shifting for voltage domain crossing. The scan generator is executed in 40-nm planar and 7-nm FINFET technologies and the 40-nm design is fabricated and verified in silicon, as shown in Figure \ref{fig7}(a). Another example is a current DAC (digital-to-analog converter) generator (Figure \ref{fig7}(b)) with programmable number of bits. Various designs with different bit numbers and technology nodes are generated and evaluated, of which simulation results are plotted in Figure \ref{fig7}(b).

\begin{table}
  \caption{Area comparisons of various custom layouts in 40-nm planar and 7-nm FinFET technologies}
  \centering
  \label{tab:matrix}
  \begin{tabular}{cccl}
    \toprule
    Type & 40nm & 7nm \\
    \midrule
    6-bit current DAC & $2,400um^2$ & $800um^2$ \\
    8-bit current DAC & $7,800um^2$ & $2,100um^2$ \\
    Custom scan & $92,400um^2$ & $20,300um^2$ \\
    & (1024bits) & (968bits) \\
  \bottomrule
\end{tabular}
\end{table}
 
\section{Conclusions}
In this paper, an automatic layout generation framework, is introduced to enhance the layout design productivity in advanced CMOS technologies. The proposed framework dynamically generates templates and grids for target technology and design flavors to produce optimal layout structures. Various post-processing features such as cut pattern generations are implemented to fulfill process-specific requirements to enhance process coverage. Advanced cyclic indexing and conditional operators are used for grid computations. Various layout structures generated from various commercial as well as open technologies reveal the effectiveness of the proposed layout generation framework.

\bibliographystyle{unsrtnat}
\bibliography{references}

\end{document}